\documentclass[12pt]{article}

\begin{document}

\noindent {\small USC-98/HEP-B5\hfill \hfill hep-th/9810025}\newline
{\small \hfill }

{\vskip 0.3cm}

\begin{center}
{\Large Hidden Symmetries, AdS}$_D\times ${\Large S}$^n${\Large , and the
Lifting of }\medskip

{\Large One-Time-Physics to Two-Time-Physics} \footnote{%
This research was partially supported by the US. Department of Energy under
grant number DE-FG03-84ER40168.}\\[0pt]

{\vskip 0.3cm}

{ITZHAK BARS}

{\vskip 0.2cm}

{Department of Physics and Astronomy, University of Southern California}

{\ Los Angeles, CA 90089-0484, USA}

{\vskip 0.3cm}

\textbf{Abstract}
\end{center}

{\small The massive non-relativistic free particle in d-1 space dimensions,
with a Lagrangian }$L=\frac m2\mathbf{\dot{r}}^2${\small ,\ has an action
with a surprising non-linearly realized }$SO(d,2)${\small \ symmetry. This
is the simplest example of a host of diverse one-time-physics systems with
hidden }$SO(d,2)${\small \ symmetric actions. By the addition of gauge
degrees of freedom, they can all be lifted to the \textit{same} }$SO(d,2)$%
{\small \ covariant unified theory that includes an extra spacelike and an
extra timelike dimension. The resulting action in d+2 dimensions has
manifest }$SO(d,2)${\small \ Lorentz symmetry and a gauge symmetry }$Sp(2,R)$%
{\small . The symmetric action defines two-time-physics. Conversely, the
two-time action can be gauge fixed to diverse one-time physical systems. In
this paper three new gauge fixed forms that correspond to the
non-relativistic particle, the massive relativistic particle, and the
particle in }$AdS_{d-n}\times S^n${\small \ curved spacetime will be
discussed at the classical level. The last case is discussed at the first
quantized and field theory levels as well. For the last case the popularly
known symmetry is }$SO(d-n-1,2)\times SO(n+1)${\small , but yet we show that
the classical or quantum versions are symmetric under the larger }$SO(d,2)$%
{\small . In the field theory version the action is symmetric under the full 
}$SO(d,2)${\small \ provided it is improved with a quantized mass term that
arises as an anomaly from operator ordering ambiguities. The anomalous mass
term vanishes for }$AdS_2\times S^0${\small \ and }$AdS_n\times S^n${\small %
\ (i.e. }$d=2n${\small ). A quantum test for the presence of
two-time-physics in a one-time-physics system is that the }$SO(d,2)${\small 
\ Casimir operators have fixed eigenvalues independent of the system. It is
shown that this test is successful for the particle in }$AdS_{d-n}\times S^n$%
{\small \ by computing the Casimirs and showing explicitly that they are
independent of }$n${\small . The strikingly larger symmetry could be
significant in the context of the proposed AdS/CFT duality. }

\vfill\eject 

\section{Hidden SO$\left( d,2\right) $ in one-time physics}

In this section we will begin by showing some examples of surprising
non-linearly realized hidden SO$\left( d,2\right) $ symmetry in simple
one-time-physics systems. We will then explain the true and systematic
origin of these symmetries, not only in these examples but also in a host of
many others, as being a simple and direct consequence of two time physics.
Two-time physics has been defined and explained in \cite{dualconf}-\cite
{dualicmp} and it will be briefly outlined below, but the reader can
understand the symmetries discussed here from the traditional
one-time-physics point of view. The main point of the examples is that the
hidden symmetry allows us to embed standard one-time-physics in a larger
spacetime with one more spacelike and one more timelike dimensions as
compared to standard one-time physics. The lifting to higher dimensions is
done with the addition of gauge degrees of freedom such that diverse actions
for one-time-physics systems converge to the \textit{same} unified action in
two-time physics that also has an Sp$\left( 2,R\right) $ gauge symmetry. The
Sp$\left( 2,R\right) $ acts on position and momentum $\left( X^M,P^M\right) $
as a doublet. This establishes an Sp$\left( 2,R\right) $ duality symmetry
among the diverse one-time-physics systems. There are consequences and some
tests of two-time-physics as will be illustrated in section 2.

\subsection{Non-relativistic particle}

\subsubsection{Hidden symmetry}

Consider the free massive non-relativistic particle in $d-1$ space
dimensions with the action 
\begin{equation}
S=\int d\tau \frac 12m\mathbf{\dot{r}}^2.  \label{nonrelaction}
\end{equation}
We will discuss this simple example from different angles because it serves
as a prototype for understanding the more complicated cases. The case of the
massless relativistic particle (with and without spin) discussed in \cite
{dualconf},\cite{dualsusy} can also serve as a prototype, but it is perhaps
not sufficiently complicated to illustrate some of the issues.

As is well known, the obvious symmetry of this system is described by the
Galilean group consisting of rotations SO$\left( d-1\right) $ and
translations $T_{d-1}$ in $\left( d-1\right) $ dimensions. The Hamiltonian $%
H=\mathbf{p}^2/2m$ commutes with the generators of these symmetries. Until
now there has not been any clue that this system has a higher symmetry
structure. However, it can be checked that the action (not the Hamiltonian)
is symmetric under the larger symmetry SO$\left( d,2\right) $ as follows.

Define a basis for an SO$\left( d,2\right) $ vector with an index $M=\left(
+^{\prime },-^{\prime },0,i\right) $, with $i=1,2,\cdots ,\left( d-1\right) $
denoting the space coordinates as in $\mathbf{r}^i$. The parameters of SO$%
\left( d,2\right) $ form an antisymmetric matrix $\varepsilon _{MN}$ with
independent components $\varepsilon _{+^{\prime }-^{\prime }},\varepsilon
_{+^{\prime }0},\varepsilon _{-^{\prime }0},\varepsilon _{+^{\prime
}i},\varepsilon _{-^{\prime }i},\varepsilon _{0i},\varepsilon _{ij}$, where
the last $\varepsilon _{ij}$ are the parameters for rotations for the
linearly realized rotations SO$\left( d-1\right) $. The hidden SO$\left(
d,2\right) $ symmetry of the action above is obtained by the following 
\textit{off-shell} linear and non-linear transformations of $\mathbf{r}%
\left( \tau \right) $%
\begin{eqnarray}
\delta \mathbf{r}^i\left( \tau \right)  &=&\varepsilon _{ij}\mathbf{\mathbf{r%
}}^j\mathbf{+}\varepsilon _{+^{\prime }-^{\prime }}\left( \mathbf{r}^i-2\tau 
\mathbf{\dot{r}}^i\right) \mathbf{+}\varepsilon _{+^{\prime }0}\frac{-\tau
\left( \mathbf{r}^i\mathbf{-\dot{r}}^i\tau \right) }{\sqrt{\left( \mathbf{r}%
-\tau \mathbf{\dot{r}}\right) ^2}}  \label{dr} \\
&&\mathbf{+}\varepsilon _{-^{\prime }0}\left[ \mathbf{\dot{r}}^i\sqrt{\left( 
\mathbf{r}-\tau \mathbf{\dot{r}}\right) ^2}-\frac{\tau \mathbf{\dot{r}}^2}2%
\frac{\left( \mathbf{r}^i\mathbf{-\dot{r}}^i\tau \right) }{\sqrt{\left( 
\mathbf{r}-\tau \mathbf{\dot{r}}\right) ^2}}\right] \mathbf{-}\varepsilon
_{+^{\prime }i}\,\tau   \nonumber \\
&&+\varepsilon _{-^{\prime }j}\left[ -\mathbf{r}^i\mathbf{\dot{r}}^j+\mathbf{%
r}^j\mathbf{\dot{r}}^i-\mathbf{r\cdot \dot{r}}\delta ^{ij}+\tau \mathbf{\dot{%
r}}^i\mathbf{\dot{r}}^j+\tau \frac{\mathbf{\dot{r}}^2}2\delta ^{ij}\right]  
\nonumber \\
&&+\varepsilon _{0j}\left[ -\delta ^{ij}\sqrt{\left( \mathbf{r}-\tau \mathbf{%
\dot{r}}\right) ^2}+\frac{\tau \mathbf{\dot{r}}^j\left( \mathbf{r}^i\mathbf{-%
\dot{r}}^i\tau \right) }{\sqrt{\left( \mathbf{r}-\tau \mathbf{\dot{r}}%
\right) ^2}}\right] .  \nonumber
\end{eqnarray}
Note the explicit $\tau $, in addition to the implicit $\tau $ in $\mathbf{r}%
^i\left( \tau \right) $, which will be related below to a gauge
transformation. The Lagrangian transforms into a total derivative $\delta
\left( \mathbf{\dot{r}}^2/2\right) =\partial _\tau \Lambda \left( \tau
,\varepsilon _{MN}\right) $, with $\Lambda \left( \tau ,\varepsilon
_{MN}\right) $ given by 
\begin{eqnarray}
\Lambda \left( \tau ,\varepsilon _{MN}\right)  &=&-\varepsilon _{+^{\prime
}-^{\prime }}\tau \mathbf{\dot{r}}^2\mathbf{-}\varepsilon _{+^{\prime
}0}\left[ \tau \frac{\mathbf{\dot{r}\cdot }\left( \mathbf{r}-\tau \mathbf{%
\dot{r}}\right) }{\sqrt{\left( \mathbf{r}-\tau \mathbf{\dot{r}}\right) ^2}}%
+\frac 12\sqrt{\left( \mathbf{r}-\tau \mathbf{\dot{r}}\right) ^2}\right] 
\label{lam} \\
&&\mathbf{+}\varepsilon _{-^{\prime }0}\left[ \frac{\mathbf{\dot{r}}^2}2%
\sqrt{\left( \mathbf{r}-\tau \mathbf{\dot{r}}\right) ^2}-\frac{\tau \mathbf{%
\dot{r}}^2}2\frac{\mathbf{\dot{r}\cdot }\left( \mathbf{r-}\tau \mathbf{\dot{r%
}}\right) }{\sqrt{\left( \mathbf{r}-\tau \mathbf{\dot{r}}\right) ^2}}\right] 
\nonumber \\
&&\mathbf{-}\varepsilon _{+^{\prime }i}\mathbf{r}^i+\varepsilon _{-^{\prime
}j}\left[ -\mathbf{\dot{r}}^j\mathbf{\dot{r}\cdot r+r}^j\frac{\mathbf{\dot{r}%
}^2}2+\tau \mathbf{\dot{r}}^j\mathbf{\dot{r}}^2\right]   \nonumber \\
&&+\varepsilon _{0j}\left[ \tau \mathbf{\dot{r}}^j\frac{\mathbf{\dot{r}\cdot 
}\left( \mathbf{r-}\tau \mathbf{\dot{r}}\right) }{\sqrt{\left( \mathbf{r}%
-\tau \mathbf{\dot{r}}\right) ^2}}\right] .  \nonumber
\end{eqnarray}
Hence the action is symmetric under SO$\left( d,2\right) .$

\subsubsection{Generators}

The generators of this SO$\left( d,2\right) $ symmetry can be derived by
using a generalized Noether theorem. Using canonical variables $\mathbf{r}%
\left( \tau \right) ,\mathbf{p}\left( \tau \right) =m\mathbf{\dot{r}}\left(
\tau \right) $ they are given at any $\tau $ by 
\begin{eqnarray}
SO\left( d-1\right)  &:&L^{ij}=\mathbf{r}^i\mathbf{p}^j-\mathbf{r}^j\mathbf{p%
}^i\quad \quad  \\
SO\left( 1,2\right)  &:&\left\{ 
\begin{array}{l}
L^{+^{\prime }-^{\prime }}=-\left( \mathbf{r}-\tau \frac{\mathbf{p}}m\right) 
\mathbf{\cdot p},\quad L^{+^{\prime }0}=-m\sqrt{\left( \mathbf{r}-\tau \frac{%
\mathbf{p}}m\right) ^2}, \\ 
L^{-^{\prime }0}=-\frac{\mathbf{p}^2}{2m}\sqrt{\left( \mathbf{r}-\tau \frac{%
\mathbf{p}}m\right) ^2},
\end{array}
\right. \quad  \\
L^{+^{\prime }i} &=&-m\left( \mathbf{r}^i-\tau \frac{\mathbf{p}^i}m\right)
,\quad \quad L^{0i}=\mathbf{p}^i\sqrt{\left( \mathbf{r}-\tau \frac{\mathbf{p}%
}m\right) ^2} \\
L^{-^{\prime }i} &=&-\frac{\mathbf{p}^2}{2m}\left( \mathbf{r}^i-\tau \frac{%
\mathbf{p}^i}m\right) +\mathbf{p}\cdot \left( \mathbf{r}-\tau \frac{\mathbf{p%
}}m\right) \frac{\mathbf{p}^i}m.
\end{eqnarray}
The Poisson brackets of these $L^{MN}\left( \tau \right) ${\small \ } form
the SO$\left( d,2\right) $ algebra at every $\tau $ (which is treated as a
parameter) 
\begin{equation}
\left\{ L^{MN},L^{RS}\right\} =\eta ^{MR}L^{NS}+\eta ^{NS}L^{MR}-\eta
^{NR}L^{MS}-\eta ^{MS}L^{NR},
\end{equation}
including the SO$\left( 1,2\right) $ and SO$\left( d-1\right) $ subalgebras
as indicated. Furthermore, the $L^{ij}$ together with $\mathbf{p}^i\sim
L^{0i}/L^{+^{\prime }0}$ form the Galilean subalgebra, which is the familiar
symmetry of the non-relativistic particle. The Galilean generators are the
only ones that do not have explicit $\tau $ dependence. The general $\tau $
dependent $L^{MN}$ generate the new hidden SO$\left( d,2\right) $ symmetries
of the action (\ref{nonrelaction}). The $\tau $ dependent terms may be
regarded as generating $\tau $-dependent local transformation on the
independent \textit{off-shell} dynamical variables $\mathbf{r}\left( \tau
\right) ,\mathbf{p}\left( \tau \right) $.

The SO$\left( d,2\right) $ transformations of the independent canonical
degrees of freedom $\mathbf{r,p}$ are obtained at any $\tau $ by evaluating
the Poisson brackets while treating $\tau $ as a parameter 
\begin{equation}
\delta \mathbf{r}^i\left( \tau \right) \mathbf{=}\frac 12\varepsilon
_{MN}\left\{ L^{MN}\left( \tau \right) ,\mathbf{r}^i\left( \tau \right)
\right\} ,\quad \delta \mathbf{p}^i\left( \tau \right) \mathbf{=}\frac
12\varepsilon _{MN}\left\{ L^{MN}\left( \tau \right) ,\mathbf{p}^i\left(
\tau \right) \right\} .  \label{drdp}
\end{equation}
Under these transformations the first order form of the action 
\begin{equation}
S=\int_0^Td\tau \left( \mathbf{\dot{r}\cdot p-}\frac{\mathbf{p}^2}{2m}%
\right) ,  \label{freenonrel}
\end{equation}
is invariant under SO$\left( d,2\right) $. Here $\mathbf{r}\left( \tau
\right) \mathbf{,p}\left( \tau \right) $ are treated as independent \textit{%
off-shell} fields whose $\tau $ dependence are unrelated to each other.
However, if they are related to each other by using the equation of motion
for momentum $\mathbf{p=}m\mathbf{\dot{r}}$, then the $\delta \mathbf{r}^i$
of (\ref{drdp}) reduces to the $\delta \mathbf{r}^i$ in (\ref{dr}) which
corresponds to the transformation law for the invariance of the action (\ref
{nonrelaction}) in the second order form.

It can be checked that the SO$\left( d,2\right) \,$generators can be
rewritten formally as the antisymmetric product of two $d+2$ dimensional
vectors in the form 
\begin{equation}
L^{MN}=X_0^MP_0^N-X_0^NP_0^M  \label{lmn0nrel}
\end{equation}
with 
\begin{eqnarray}
M &=&\left( +^{\prime }\quad ,\quad -^{\prime }\quad \quad ,\quad \quad
0\quad \quad \quad ,\quad \quad \quad i\quad \right) \\
X_0^M &=&\left( \tau ,\,\,\,\frac{\mathbf{r\cdot p}}m\mathbf{-}\frac{\tau 
\mathbf{p}^2}{2m^2},\,\,\sqrt{\left( \mathbf{r}-\tau \frac{\mathbf{p}}%
m\right) ^2},\quad \mathbf{r}^i\right)  \label{nrelX} \\
P_0^M &=&\left( m,\quad \quad \frac{\mathbf{p}^2}{2m}\quad ,\quad \quad
0\quad \quad ,\quad \quad \quad \mathbf{p}^i\right) .  \label{nrelP}
\end{eqnarray}
These satisfy $X_0^2=P_0^2=X_0\cdot P_0=0$ with a metric $\eta _{MN}$, such
that $\eta _{+^{\prime }-^{\prime }}=-1,\,\eta _{00}=-1$, $\eta _{ij}=\delta
_{ij}$. This is the metric invariant under SO$\left( d,2\right) $ with two
timelike dimensions.

\subsubsection{Lifting to two-time-physics}

The SO$\left( d,2\right) $ symmetry with this structure implies that the
non-relativistic particle action can be lifted to a manifestly SO$\left(
d,2\right) $ symmetric form by the addition of gauge degrees of freedom.
From the form of (\ref{lmn0nrel}) we can deduce that the manifestly
symmetric form of the symmetry is the Lorentz symmetry SO$\left( d,2\right) $
realized linearly on a vector $X^M\left( \tau \right) $ and its canonical
conjugate $P^M\left( \tau \right) $. These describe a particle (0-brane) in
a spacetime with $d$ spacelike and 2 timelike dimensions ($X^M,P^M$ are
lifted forms of $X_0^M,P_0^M$ including gauge degrees of freedom). This
shows that the non-relativistic particle is connected to the realm of
two-time-physics, a formulation that also has a sufficiently large gauge
symmetry Sp$\left( 2,R\right) $ to kill all ghosts and connect back to
one-time-physics as discussed in recent papers \cite{dualconf}-\cite
{dualicmp}.

The Sp$\left( 2,R\right) $ gauge theory for zero branes takes the form \cite
{dualconf} 
\begin{eqnarray}
S_0 &=&\frac 12\int_0^Td\tau \left( D_\tau X_i^M\right) \varepsilon
^{ij}X_j^N\eta _{MN}  \label{action} \\
&=&\int_0^Td\tau \left( \partial _\tau X_1^MX_2^N-\frac
12A^{ij}X_i^MX_j^N\right) \eta _{MN}\,\,.  \nonumber
\end{eqnarray}
The canonical conjugates are $X_1^M=X^M$ and $\partial S/\partial \dot{X}%
_1^M=X_2^M=P^M$. They are consistent with the idea that $(X_1^M,X_2^M)$ is
the Sp$\left( 2,R\right) $ doublet $\left( X^M,P^M\right) $. The symmetric $%
A^{ij}$ are the 3 gauge potentials of Sp$\left( 2,R\right) $. The equations
of motion for $A^{ij}$ give the first class constraints 
\begin{equation}
X\cdot X=X\cdot P=P\cdot P=0  \label{constraints}
\end{equation}
that form the Sp$\left( 2,R\right) $ Lie algebra. The action is evidently
symmetric under SO$\left( d,2\right) $. The generators are gauge invariant 
\begin{equation}
L^{MN}=X_i^MX_j^N\varepsilon ^{ij}=X^MP^N-X^NP^M.  \label{lmn}
\end{equation}
In this form all components of $X^M$ and $P^M$ are canonical and $\delta
X^M,\delta P^M$ are obtained by using the basic Poisson brackets $\delta
X^M=\frac 12\varepsilon _{RS}\left\{ L^{RS},X^M\right\} $, etc.. In this
fully covariant approach the constraints are applied on the states, as
discussed in \cite{dualconf}-\cite{dualicmp}$.$

The three gauge choices that reduce the general system to the
non-relativistic particle are 
\begin{equation}
X^{+^{\prime }}\left( \tau \right) =\tau ,\quad P^{+^{\prime }}\left( \tau
\right) =m,\quad P^0\left( \tau \right) =0.  \label{nrelgauge}
\end{equation}
After solving the three constraints (\ref{constraints}) explicitly in this
gauge, $X^M\left( \tau \right) $ and $P^M\left( \tau \right) $ take the form
given in (\ref{nrelX},\ref{nrelP}) . Note that the non-relativistic particle
action (\ref{freenonrel}) can then be written as 
\begin{equation}
S=\int d\tau \,\,\partial _\tau X_0\cdot P_0=\int_0^Td\tau \left( \mathbf{%
\dot{r}\cdot p-}\frac{\mathbf{p}^2}{2m}\right) .
\end{equation}
This follows from the fully gauge invariant and SO$\left( d,2\right) $
invariant two-time-physics action (\ref{action}) after the gauge (\ref{nrelX}%
,\ref{nrelP}) has been inserted.

\subsubsection{An intermediate gauge}

It is also interesting to consider an intermediate gauge. For example, if we
choose only two gauges $P^{+^{\prime }}\left( \tau \right) =m,$ $P^0\left(
\tau \right) =0$ and solve two constraints $X^2=X\cdot P=0$, there remains
one gauge freedom and one constraint. Then $X^M,P^M$ are parametrized in
terms of the $d$ canonical degrees of freedom $\left( t\left( \tau \right) ,%
\mathbf{r}^i\left( \tau \right) \right) $ and their canonical conjugates $%
\left( H\left( \tau \right) ,\mathbf{p}^i\left( \tau \right) \right) $ as
follows 
\begin{eqnarray}
M &=&\left( +^{\prime }\quad ,\quad -^{\prime }\quad \quad ,\quad \quad
0\quad \quad \quad ,\quad \quad \quad i\quad \right)  \nonumber \\
X^M &=&\left( t,\,\,\,\frac{\mathbf{r\cdot p}}m\mathbf{-}t\frac Hm,\,\,\sqrt{%
\mathbf{r}^2-2\frac{t\mathbf{r\cdot p}}m\mathbf{+}2\frac Hmt^2},\quad 
\mathbf{r}^i\right)  \label{nrcX} \\
P^M &=&\left( m,\quad \quad H\quad ,\quad \quad 0\quad \quad ,\quad \quad
\quad \mathbf{p}^i\right) \,\quad  \label{nrcP}
\end{eqnarray}
We derive the dynamics for the remaining degrees of freedom $t,\mathbf{r,}H%
\mathbf{,p}$ by inserting this gauge fixed form in the original action (\ref
{action}). The result is a one-time action given by 
\begin{eqnarray}
S &=&\int_0^Td\tau \left( \partial _\tau X^MP^N-\frac 12A^{22}\left( -2mH+%
\mathbf{p}^2\right) -0-0\right) \\
&=&\int_0^Td\tau \left[ -H\partial _\tau t+\mathbf{p}^i\partial _\tau 
\mathbf{r}^i-\frac 12A^{22}\left( -2mH+\mathbf{p}^2\right) \right] .
\label{nrtau}
\end{eqnarray}
We have dropped a total derivative term $\partial _\tau \left( \mathbf{%
r\cdot p}\right) $ that does not contribute to the dynamics. The last form
of the action confirms that $\left( t,H\right) $ and $\left( \mathbf{r,p}%
\right) $ are canonical conjugates with Poisson brackets 
\begin{equation}
\left\{ t,H\right\} =-1,\quad \left\{ \mathbf{r}^i,\mathbf{p}^j\right\}
=\delta ^{ij}.  \label{nrcommutators}
\end{equation}
The $A^{22}$ equation of motion gives the constraint $H=\frac{\mathbf{p}^2}{%
2m}$. This is the same as the $P^2=0$ constraint. The remaining local
symmetry corresponds to $\tau $ reparametrizations. In the gauge $t\left(
\tau \right) =\tau $ the dynamics describes the free non-relativistic
massive particle. In fact, if this additional gauge is chosen the action
reduces to (\ref{freenonrel}).

We expect that this form of one-time-physics action (\ref{nrtau}) is also
symmetric under SO$\left( d,2\right) $. To construct the generators we
insert the gauge choice of (\ref{nrcX},\ref{nrcP}) in the gauge invariant $%
L^{MN}$ of (\ref{lmn}). At the classical level, without watching orders of
operators, they are given by (now there is no explicit $\tau $ dependence) 
\begin{eqnarray}
SO\left( d-1\right) &:&L^{ij}=\mathbf{r}^i\mathbf{p}^j-\mathbf{r}^j\mathbf{p}%
^i  \label{nrlmn1} \\
SO\left( 1,2\right) &:&\left\{ 
\begin{array}{l}
L^{+^{\prime }-^{\prime }}=2tH-\mathbf{r\cdot p},\quad L^{+^{\prime }0}=-m%
\sqrt{\mathbf{r}^2-2t\frac{\mathbf{r\cdot p}}m\mathbf{+}2\frac Hmt^2}, \\ 
L^{-^{\prime }0}=-H\sqrt{\mathbf{r}^2-2t\frac{\mathbf{r\cdot p}}m\mathbf{+}%
2\frac Hmt^2},
\end{array}
\right. \quad \\
L^{+^{\prime }i} &=&t\mathbf{p}^i-m\mathbf{r}^i,\quad L^{-^{\prime }i}=%
\mathbf{-}t\frac Hm\mathbf{p}^i+\frac{\mathbf{r\cdot p}}m\mathbf{p}^i-H%
\mathbf{r}^i, \\
L^{0i} &=&\mathbf{p}^i\sqrt{\mathbf{r}^2-2\frac{t\mathbf{r\cdot p}}m\mathbf{+%
}2\frac Hmt^2}.  \label{nrlmn4}
\end{eqnarray}
Using the basic Poisson brackets (\ref{nrcommutators}) it can be shown that
these $L^{MN}$ form the SO$\left( d,2\right) $ algebra. They also generate
the transformation rules for $t,H,\mathbf{r}^i,\mathbf{p}^i$ by evaluating
the Poisson brackets $\delta t=\frac 12\varepsilon _{MN}\left\{
L^{MN},t\right\} $, etc.. The action is not invariant under these
transformations alone; for invariance one must also transform $A^{22}$. The
reason is that the constraint $\left( -2mH+\mathbf{p}^2\right) $ that
multiplies $A^{22}$ in the action is not invariant, but transforms into
itself with an overall factor 
\begin{equation}
\delta \left( -2mH+\mathbf{p}^2\right) =\gamma \left( \varepsilon _{MN},\tau
\right) \times \left( -2mH+\mathbf{p}^2\right) ,
\end{equation}
where 
\begin{equation}
\gamma \left( \varepsilon _{MN},\tau \right) =\left( \frac{\mathbf{p}%
^j\left( \tau \right) }m\varepsilon _{-^{\prime }j}-\varepsilon _{+^{\prime
}-^{\prime }}+\frac{t\left( \tau \right) \left( \varepsilon _{+^{\prime }0}+%
\frac{H\left( \tau \right) }m\varepsilon _{-^{\prime }0}-\varepsilon _{0j}%
\frac{\mathbf{p}^j\left( \tau \right) }m\right) }{\sqrt{\mathbf{r}^2\left(
\tau \right) -2t\left( \tau \right) \frac{\mathbf{r}\left( \tau \right) 
\mathbf{\cdot p}\left( \tau \right) }m\mathbf{+}2\frac{H\mathbf{\left( \tau
\right) }}mt^2\left( \tau \right) }}\right) .
\end{equation}
This term is cancelled by taking $\delta A^{22}=-2A^{22}\gamma \left(
\varepsilon _{MN},\tau \right) $. This factor can be understood as follows.
Recall that when a gauge is fixed the new generators $L^{MN}$ perform a
naive SO$\left( d,2\right) $ transformation (that disturbs the gauge)
followed by an Sp$\left( 2,R\right) $ gauge transformation (that restores
the gauge). The constraints (\ref{constraints}) transform as a triplet under
the restoring gauge transformation.. Since two of the constraints are
already zero explicitly, the third one transforms into itself with an
overall factor $\delta \left( P^2\right) =\gamma \left( \varepsilon
_{MN}\right) \times P^2$, and this must be compensated by the transformation
of the gauge field $\delta A^{22}$ as given above.

\subsubsection{Field theory}

When we do not make the last gauge choice the remaining constraint must be
applied on the states. A complete Hilbert space for the quantum theory is
given in configuration space as $|t,\mathbf{r>}$. The physical subset of
states $|\psi >$ are those that satisfy the constraint 
\begin{equation}
\left( H-\frac{\mathbf{p}^2}{2m}\right) |\psi >=0.
\end{equation}
In terms of the wavefunction in configuration space $\psi \left( t,\mathbf{r}%
\right) =<t,\mathbf{r}|\psi >$ the physical state condition takes the form
of the non-relativistic Schr\"{o}dinger equation 
\begin{equation}
i\partial _t\psi \left( t,\mathbf{r}\right) =-\frac{\mathbf{\nabla }^2}{2m}%
\psi \left( t,\mathbf{r}\right) .  \label{schreq}
\end{equation}
The effective field theory that reproduces this equation is 
\begin{equation}
S_{eff}=\int dtd\mathbf{r\,}\left[ i\psi ^{*}\partial _t\psi -\frac 1{2m}%
\mathbf{\nabla }\psi ^{*}\cdot \mathbf{\nabla }\psi \right] .
\label{schraction}
\end{equation}
The norm of the physical state is then given by integrating the time
component of the probability current at fixed time 
\begin{equation}
<\psi |\psi >=\int d^{d-1}\mathbf{r\,\,}\psi ^{*}\left( t,\mathbf{r}\right)
\psi \left( t,\mathbf{r}\right)
\end{equation}
This norm is independent of $t$ due to the conservation of the probability
current $\left( \psi ^{*}\psi ,\frac{\psi ^{*}\mathbf{\nabla }\psi -\mathbf{%
\nabla }\psi ^{*}\psi }{2im}\right) $ as a result of the physical state
condition (\ref{schreq}).

Now we ask the question: is the field theoretic version of the theory also SO%
$\left( d,2\right) $ invariant under the transformation 
\begin{equation}
\delta \psi \left( t,\mathbf{r}\right) =\frac i2\varepsilon _{MN}<t,\mathbf{%
r|}L^{MN}|\psi >=\frac 12\varepsilon _{MN}\left( \hat{L}^{MN}\psi \right)
\left( t,\mathbf{r}\right) \,\,
\end{equation}
where $\hat{L}^{MN}$ are differential operators obtained from the operators $%
L^{MN}$ in (\ref{nrlmn1}-\ref{nrlmn4}) by replacing $H=i\hbar \partial _t$,
and $\,\mathbf{p=-}i\hbar \mathbf{\nabla }$ as applied on $\psi \left( t,%
\mathbf{r}\right) $. The correct quantum operators to all orders of $\hbar $
must correspond to a particular order of the canonical operators $t,H,%
\mathbf{r,p}$ , but we have not attempted to find the order. Here we face a
difficult problem with the non-linear form of the $L^{MN}$ since an infinite
number of possibilities of ordering of an infinite series is possible.
Therefore we have not been able to give a definitive answer to this question%
\footnote{%
The analogous question for the AdS$_{d-n}\times S^n$ will be answered in the
affirmative in the last section.}. It would be amazing if one can find an
ordering of operators that would give SO$\left( d,2\right) $ invariance for
the non-relativistic Schr\"{o}dinger field theory action (\ref{schraction}).
If there is no such order, it would imply that the quantum theory in the
form (\ref{schraction}) produces anomalies that break the SO$\left(
d,2\right) $ symmetry. If this is the case one may ask if there is an
anomalous term that can be added to the field theory to yield the correct
quantum version with an SO$\left( d,2\right) $ symmetry. This question
remains open for now .

\subsection{Massive relativistic particle}

\subsubsection{Lifting to the intermediate SO$\left( d-1,1\right) $
covariant gauge}

To understand better the hidden symmetries and their origins it is useful to
start with the fully gauge fixed form of the relativistic massive particle
action and first lift it to the intermediate gauge which is manifestly SO$%
\left( d-1,1\right) $ Lorentz covariant. The answer is well known, but by
using similar steps as the previous section it may be helpful to make
analogies to the non-relativistic case, thus clarifying some concepts that
may have remained hazy to the reader. Consider the action for the massive
relativistic particle 
\begin{equation}
S=m\int_0^Td\tau \sqrt{1-\mathbf{\dot{r}}^2},  \label{relaction}
\end{equation}
which as (\ref{nonrelaction}) is also symmetric under rotations and
translations. This action has a ``hidden'' off-shell symmetry under $\delta 
\mathbf{r}\left( \tau \right) =\mathbf{\beta }^i\tau -\mathbf{\beta \cdot r}%
\left( \tau \right) \mathbf{\,\dot{r}\,}^i\left( \tau \right) $, where $%
\mathbf{\beta }^i$ are constant parameters, since the Lagrangian transforms
into a total derivative $\delta \sqrt{1-\mathbf{\dot{r}}^2}=\partial _\tau
\left[ \mathbf{r\cdot \beta }\sqrt{1-\mathbf{\dot{r}}^2}\right] $. Using a
generalized Noether's theorem one can derive the generator of this
transformation, and by writing it in terms of the canonical variables $%
\mathbf{r}\left( \tau \right) ,\mathbf{p}\left( \tau \right) =m\mathbf{\dot{r%
}/}\sqrt{1-\mathbf{\dot{r}}^2}$ in the form 
\begin{equation}
\mathbf{K}^i\left( \tau \right) =\tau \mathbf{p}^i\left( \tau \right) -%
\mathbf{r}^i\left( \tau \right) \sqrt{\mathbf{p}^2\left( \tau \right) +m^2,}
\label{ki}
\end{equation}
one can recognize that it is the generator of relativistic boosts. The $%
\delta \mathbf{r}\left( \tau \right) $ used above can be written as the
Poisson bracket $\delta \mathbf{r}\left( \tau \right) =\left\{ -\mathbf{%
\beta \cdot K}\left( \tau \right) ,\,\mathbf{r}^i\left( \tau \right)
\right\} $. Note the explicit $\tau $ dependence in $\mathbf{K}^i\left( \tau
\right) $ and in $\delta \mathbf{r}^i\left( \tau \right) $ which is
analogous to the explicit $\tau $ that appeared in the previous
non-relativistic case. Although the action is symmetric under the boosts,
the Hamiltonian $H=\sqrt{\mathbf{p}^2+m^2}$ is not symmetric, but transforms
under them in a well defined manner. We can compare the ``hidden'' boost
symmetry of (\ref{relaction}) to a subset of the hidden symmetries SO$\left(
d,2\right) $ of (\ref{nonrelaction}).

The ``hidden'' boost symmetry can be made manifest by lifting the action (%
\ref{relaction}) to its well known Lorentz symmetric form 
\begin{equation}
S=m\int d\tau \sqrt{-\left( \dot{x}^\mu \right) ^2}.  \label{relact}
\end{equation}
To do this lifting we must add gauge degrees of freedom and then the action
is gauge invariant under $\tau $-reparametrizations. As is well known this
action can be rewritten in the first order form by introducing the canonical
momentum $p^\mu \left( \tau \right) $ and an einbein $A^{22}\left( \tau
\right) $ that plays the role of a Lagrange multiplier to implement the
constraint on this momentum 
\begin{equation}
S=\int_0^Td\tau \left[ \dot{x}^\mu p_\mu -\frac 12A^{22}\left( p_\mu
^2+m^2\right) \right] .  \label{relact2}
\end{equation}
Integrating out $p^\mu $ and $A^{22}$ gives back (\ref{relact}). This form
should be compared to the non-relativistic case in eq.(\ref{nrtau}). The
generators of the Lorentz symmetry are now given in terms of canonical
variables $L^{\mu \nu }=x^\mu p^\nu -x^\nu p^\mu $ while the constraint is
applied on the physical states. Fixing the gauge $x^0\left( \tau \right)
=\tau $ reduces the action (\ref{relact}) back to (\ref{relaction}) while $%
L^{0i}$ becomes the $\mathbf{K}^i\left( \tau \right) $ of (\ref{ki}).

\subsubsection{Lifting to two-time-physics}

We now note the surprising SO$\left( d,2\right) $ symmetry of the action (%
\ref{relact2}) as follows. Using the basis for a $d+2$ dimensional vector
with index $M=\left( 0^{\prime },1^{\prime },\mu \right) $ the parameters of
SO$\left( d,2\right) $ are given as an antisymmetric tensor with components $%
\varepsilon _{0^{\prime }1^{\prime }},\varepsilon _{0^{\prime }\mu
},\varepsilon _{1^{\prime }\mu },\varepsilon _{\mu \nu }$. The last $%
\varepsilon _{\mu \nu }$ correspond to the linearly realized Lorentz
symmetry. The full linearly and non-linearly realized \textit{off-shell} SO$%
\left( d,2\right) $ transformation is 
\begin{eqnarray}
\delta x^\mu &=&\varepsilon ^{\mu \nu }x_\nu -\varepsilon _{0^{\prime
}1^{\prime }}\frac{x^\mu x\cdot p}{\sqrt{m^2x^2+\left( x\cdot p\right) ^2}}%
+\varepsilon _{1^{\prime }\nu }\left[ \eta ^{\nu \mu }\frac{x\cdot p}m+\frac{%
p^\nu }mx^\mu \right] \\
&&-\varepsilon _{0^{\prime }\nu }\left[ \frac{p^\nu }m\frac{x^\mu x\cdot p}{%
\sqrt{m^2x^2+\left( x\cdot p\right) ^2}}+\eta ^{\nu \mu }\frac{\sqrt{%
m^2x^2+\left( x\cdot p\right) ^2}}m\right]  \nonumber
\end{eqnarray}
and 
\begin{eqnarray}
\delta p^\mu &=&\varepsilon ^{\mu \nu }p_\nu +\varepsilon _{0^{\prime
}1^{\prime }}\frac{m^2x^\mu +x\cdot p\,p^\mu }{\sqrt{m^2x^2+\left( x\cdot
p\right) ^2}}-\varepsilon _{1^{\prime }\nu }\left[ \eta ^{\nu \mu }m+\frac{%
p^\nu }mp^\mu \right] \\
&&-\varepsilon _{0^{\prime }\nu }\,p^\nu \frac{m^2x^\mu +x\cdot p\,p^\mu }{%
\sqrt{m^2x^2+\left( x\cdot p\right) ^2}}  \nonumber
\end{eqnarray}
and 
\begin{equation}
\delta A^{22}=A^{22}\left[ \frac{\left( \varepsilon _{0^{\prime }1^{\prime
}}+\varepsilon _{0^{\prime }\nu }\frac{p^\nu }m\right) x\cdot p}{\sqrt{%
m^2x^2+\left( x\cdot p\right) ^2}}+2\varepsilon _{1^{\prime }\nu }\frac{%
p^\nu }m\right] +\varepsilon _{1^{\prime }\nu }\frac{\dot{x}^\nu }m
\end{equation}
This transformation gives a total derivative $\delta L=\partial _\tau
\Lambda \left( \varepsilon _{MN},\tau \right) $ with 
\begin{eqnarray}
\Lambda \left( \varepsilon _{MN},\tau \right) &=&\varepsilon _{0^{\prime
}1^{\prime }}\sqrt{m^2x^2+\left( x\cdot p\right) ^2}+\varepsilon _{1^{\prime
}\nu }p^\nu x\cdot p-m\varepsilon _{1^{\prime }\nu }x^\nu \\
&&+\varepsilon _{0^{\prime }\nu }\frac{p^\nu }m\frac{\left( x\cdot p\right)
^2}{\sqrt{m^2x^2+\left( x\cdot p\right) ^2}}.  \nonumber
\end{eqnarray}
Hence the action (\ref{relact2}) is invariant.

The generators of this transformation are 
\begin{eqnarray}
L^{0^{\prime }1^{\prime }} &=&\sqrt{m^2x^2+\left( x\cdot p\right) ^2},\quad
L^{0^{\prime }\mu }=p^\mu \sqrt{m^2x^2+\left( x\cdot p\right) ^2} \\
L^{1^{\prime }\mu } &=&-\frac{x\cdot p}mp^\mu -mx^\mu ,\quad L^{\mu \nu
}=x^\mu p^\nu -x^\nu p^\mu
\end{eqnarray}
They close under Poisson brackets to form the SO$\left( d,2\right) $ Lie
algebra.

These generators are written in the form of cross products 
\begin{equation}
L^{MN}=X_0^MP_0^N-X_0^NP_0^M
\end{equation}
with 
\begin{eqnarray}
M &=&\left( \quad \quad \quad 0^{\prime }\quad \quad \quad ,\quad \quad
1^{\prime }\quad \quad ,\quad \mu \right)  \nonumber \\
X_0^M &=&\left( \sqrt{x^2+\left( \frac{x\cdot p}m\right) ^2},\,\,-\frac{%
x\cdot p}m\,,\quad x^\mu \right)  \label{relgauge} \\
P_0^M &=&\left( \quad \quad \quad 0\quad \quad \quad ,\quad \quad m\quad
,\quad p^\mu \right) .  \label{relgauge2}
\end{eqnarray}
These satisfy $X_0^2=X_0\cdot P_0=0$ while $P_0^2=p^2+m^2$, with the metric $%
\eta ^{0^{\prime }0^{\prime }}=\eta ^{1^{\prime }1^{\prime }}=-\eta
^{0^{\prime }0^{\prime }}=1$ and $\eta ^{\mu \nu }\,$= Minkowski. This form
suggests that we may lift the system to two-time physics.

Therefore we may start from the Sp$\left( 2,R\right) $ gauge symmetric
two-time physics action (\ref{action}), choose the two gauges 
\begin{equation}
P^{0^{\prime }}\left( \tau \right) =0,\,\quad P^{1^{\prime }}\left( \tau
\right) =m,
\end{equation}
and solve the two constraints $X^2=X\cdot P=0$. The result is the gauge
fixed form (\ref{relgauge},\ref{relgauge2}). The dynamics of the remaining
degrees of freedom $\left( x^\mu ,p^\mu \right) $ is obtained by inserting
the gauge fixed form (\ref{relgauge},\ref{relgauge2}) into the two-time
physics action (\ref{action}). The result is the one-time-physics action (%
\ref{relact2}) for the relativistic particle. This action has one remaining
gauge symmetry ($\tau $ reparametrization) and imposes the remaining
constraint $P^2=p^2+m^2=0$ as the equation of motion for $A^{22}$.

This shows that both the relativistic and the non-relativistic particle are
lifted to the \textit{same} two-time-physics action. Hence to an observer in
two-time physics these two systems are the same, since they are just gauge
fixed versions of the same theory.

\subsection{Particle on $AdS_{d-n}\times S^n$}

A particle moving in the curved background $AdS_{d-n}\times S^n$ is
described by the action 
\begin{equation}
S=\int_0^Td\tau \,\left( G_{\mu \nu }^{AdS}\left( x\right) \,\dot{x}^\mu 
\dot{x}^\nu +G_{ab}^{S^n}\left( y\right) \,\dot{y}^a\dot{y}^b\right) .
\label{asdsnaction}
\end{equation}
where $m=0,1,\cdots ,d-n-1$ and $a=1,2,\cdots ,n$. There are many ways of
parametrizing the AdS metric. The particular parametrization used below is
convenient for discussing and resolving the quantum ordering problem which
will be dealt with in the next section. The point that we will make is that
for AdS$_D\times S^n$ the full symmetry of the action is SO$\left(
D+n,2\right) .$ Furthermore, as long as $D+n=d$ is a constant the various
models distinguished by $n$ are Sp$\left( 2,R\right) $ dual to each other
because they are obtained from the same Sp$\left( 2,R\right) $ gauge
symmetric two-time-physics action by gauge fixing.

As in the previous cases, the larger SO$\left( d,2\right) $ symmetry comes
as a surprise since the popularly known symmetry in this background is SO$%
\left( d-n-1,2\right) \times SO\left( n+1\right) $ which is smaller than SO$%
\left( d,2\right) $. For example, we claim that the action for AdS$_3$ alone
has SO$\left( 3,2\right) $ symmetry which is larger than the popularly known
SO$\left( 2,2\right) $. Similarly the action for AdS$_5\times S^5$ has SO$%
\left( 10,2\right) $ symmetry which is larger than the popularly known SO$%
\left( 4,2\right) \times SO\left( 6\right) $; and the action for AdS$%
_4\times S^7$ or AdS$_7\times S^4$ has SO$\left( 11,2\right) $ symmetry.

Instead of lifting the $AdS_{d-n}\times S^n$ action (\ref{asdsnaction}) to
the two-time-physics action (\ref{action}), we will construct (\ref
{asdsnaction}) as a gauge fixed form of (\ref{action}). Lifting would
correspond to the reverse process.

Consider the $d+2$ dimensional vectors $X^M,P^M$ in the basis $M=\left(
+^{\prime },-^{\prime },\mu ,i\right) $ for $\mu =0,1,\cdots ,d-n-1$ and $%
i=1,2,\cdots ,n+1.$ The metric is $\eta ^{+^{\prime }-^{\prime }}=-1,\,\eta
^{ij}=\delta ^{ij}$ and $\eta ^{\mu \nu }$ = Minkowski. We choose two gauges
by demanding $\left| X^i\right| =1$ and $P^{+^{\prime }}=0.$ Then the unit
vector $X^i=\frac{\mathbf{u}^i}{\left| \mathbf{u}\right| }\equiv \mathbf{%
\Omega }^i$ describes a sphere $S^n$ as the boundary of a ball in $n+1$
dimensions. The radius of the ball $\left| X^i\right| $ is one of the
coordinates that has been gauge fixed to 1. The constraints $X^2=X\cdot P=0$
are solved by the following parametrization 
\begin{eqnarray}
M &=&\left( \,\,+^{\prime }\quad \quad \,\,\,-^{\prime }\quad \quad
\,\,\,\mu \quad \quad \,i\right)   \nonumber \\
X^M &=&\left( \left| \mathbf{u}\right| ,\,\,\,\,\,\,\,\frac{1+\mathbf{u}^2x^2%
}{2\left| \mathbf{u}\right| },\quad \left| \mathbf{u}\right| x^\mu \,,\quad 
\frac{\mathbf{u}^i}{\left| \mathbf{u}\right| }\right)   \label{adsX} \\
P^M &=&\left( 0,\,\,\,-\frac{\mathbf{u\cdot k}}{\left| \mathbf{u}\right| }+%
\frac{x\cdot p}{\left| \mathbf{u}\right| }\,\,,\,\,\,\,\,\,\,\,\,\frac{p^\mu 
}{\left| \mathbf{u}\right| }\,\,\,,\,\,\mathbf{\,}\left( \left| \mathbf{u}%
\right| \mathbf{k}^i-2\mathbf{k\cdot u}\,\frac{\mathbf{u}^i}{\left| \mathbf{u%
}\right| }\right) \right)   \label{adsP}
\end{eqnarray}
The bold vectors $\mathbf{u}^i\mathbf{,k}^i$ are in $n+1$ dimensions and $%
x^\mu ,p^\mu $ are in $d-n-1$ dimensions. For $n=0$ we replace $\frac{%
\mathbf{u}^i}{\left| \mathbf{u}\right| }$ by $1$. Inserting this gauge fixed
form into the original two-time physics action (\ref{action}) gives an
action that determines the dynamics of $x^\mu \left( \tau \right) ,p^\mu
\left( \tau \right) ,\mathbf{u}^i\left( \tau \right) ,\mathbf{k}^i\left(
\tau \right) $%
\begin{eqnarray}
S &=&\int d\tau \left( p\cdot \dot{x}+\mathbf{k}\cdot \mathbf{\dot{u}}-\frac
12A^{22}\left( \frac{p^2}{\mathbf{u}^2}+\mathbf{u}^2\mathbf{\,k}^2\right)
\right)   \label{adsaction1} \\
&\longrightarrow &\int d\tau \frac 1{2A^{22}}\left( \frac{\mathbf{\dot{u}}^2%
}{\mathbf{u}^2}+\mathbf{u}^2\dot{x}^2\right)   \label{adsaction} \\
&=&\int d\tau \frac 1{2A^{22}}\left( \frac{\dot{u}^2}{u^2}+u^2\dot{x}^2+%
\mathbf{\dot{\Omega}}^2\right) 
\end{eqnarray}
The second form of the action is obtained by integrating out the momenta.
From the first line we see that the vectors $p^\mu ,\mathbf{k}^i$ are indeed
the canonical conjugates to $x^\mu ,\mathbf{u}^i$ respectively. The last
line is obtained by making a transformation from Cartesian coordinates to
spherical coordinates $\mathbf{u}^i=u\mathbf{\Omega }^i$. This action
describes the particle in the curved background $AdS_{d-n}\times S^n$ with
metric 
\[
ds^2=u^2\left( dx^\mu \right) ^2+\frac{\left( du\right) ^2}{u^2}+\left( d%
\mathbf{\Omega }\right) ^2
\]
where the $d-n$ coordinates of AdS$_{d-n}$ are $\left( x^\mu ,u\right) $ and
the $n$ coordinates of $S^n$ are those that parametrize the unit vector $%
\mathbf{\Omega }^i$ embedded in $n+1$ dimensions. This form of the metric
has been used in recent discussions of the proposed AdS-CFT duality \cite
{maldacena}, and we find it useful for the discussion of operator ordering
that will be dealt with in the next section\footnote{%
There is a similarity between our parametrization and one used in \cite
{kalltown}, however our's treats the radius of AdS or the sphere (here
scaled to $R=1$) as an additional coordinate that has been gauge fixed (i.e. 
$\left| X^i\left( \tau \right) \right| =R=1$). This additional coordinate
together with the global and gauge symmetries of the action (\ref{action})
is what permits us to have the larger symmetry SO$\left( d,2\right) \supset
SO\left( d-n-1,2\right) \times SO\left( n+1\right) $.}. There are many other
possible parametrizations of the AdS metric. Each one of them will
correspond to some form of gauge choice in our formalism. For such other
gauge choices for AdS see \cite{dualicmp} and \cite{dualsusy}.

The point here is that our construction shows that the symmetry of the
action is SO$\left( d,2\right) $ which is larger than the popularly known SO$%
\left( d-n-1,2\right) \times SO\left( n+1\right) $. In our approach the SO$%
\left( d,2\right) $ generators are obtained by inserting the gauge fixed
forms of $X_0^M$ and $P_0^M$ given in (\ref{adsX},\ref{adsP}) into the gauge
invariant $L^{MN}$ of (\ref{lmn})$.$ At the classical level (operator
ordering ignored) we obtain $L^{MN}=X_0^MP_0^N-X_0^NP_0^M$ in the form 
\begin{eqnarray}
L^{+^{\prime }-^{\prime }} &=&-\mathbf{u\cdot k}+x\cdot p,\quad L^{+^{\prime
}\mu }=p^\mu ,\quad L^{+^{\prime }i}=\mathbf{u}^2\mathbf{k}^i\,\mathbf{\,}%
-2\,\mathbf{\,k\cdot u}\,\mathbf{u}^i  \label{lmnclass1} \\
L^{-^{\prime }\mu } &=&\frac{p^\mu }{2\mathbf{u}^2}+\mathbf{u\cdot k}x^\mu
+\frac 12x^2p^\mu -x\cdot px^\mu \\
L^{-^{\prime }i} &=&\frac 12\mathbf{k}^i+\frac{x^2}2\left( \mathbf{u}^2%
\mathbf{k}^i\,\mathbf{\,}-2\,\mathbf{\,k\cdot u}\,\mathbf{u}^i\right)
-x\cdot p\frac{\mathbf{u}^i}{\mathbf{u}^2} \\
L^{\mu \nu } &=&x^\mu p^\nu -x^\nu p^\mu ,\quad L^{ij}=\mathbf{u}^i\mathbf{k}%
^j-\mathbf{u}^j\mathbf{k}^i \\
L^{\mu i} &=&x^\mu \mathbf{\,}\left( \mathbf{u}^2\mathbf{k}^i\,\mathbf{\,}%
-2\,\mathbf{\,k\cdot u}\,\mathbf{u}^i\right) -p^\mu \frac{\mathbf{u}^i}{%
\mathbf{u}^2}  \label{lmnclass5}
\end{eqnarray}
By using the basic Poisson brackets among $\left( \mathbf{u},\mathbf{k}%
\right) ,\left( x^\mu ,p^\mu \right) $ it is easily seen that these form the
SO$\left( d,2\right) $ algebra 
\begin{equation}
\left\{ L^{MN},L^{RS}\right\} =\eta ^{MR}L^{NS}+\eta ^{NS}L^{MR}-\eta
^{NR}L^{MS}-\eta ^{MS}L^{NR}.
\end{equation}

The generators for the subgroup $SO\left( n+1\right) \times $SO$\left(
d-n-1,2\right) $ are $L^{ij}$ and $\left( L^{\mu \nu },L^{+^{\prime
}-^{\prime }},L^{+^{\prime }\mu },L^{-^{\prime }\mu }\right) $ respectively.
The additional symmetry generators that complete to SO$\left( d,2\right) $
are $L^{+^{\prime }i},L^{-^{\prime }i},L^{\mu i}$. It is well known that the
action (\ref{adsaction}) is symmetric under $SO\left( n+1\right) \times $SO$%
\left( d-n-1,2\right) $. To show that it is also symmetric under the full SO$%
\left( d,2\right) $ it is sufficient to show that it is symmetric under the $%
L^{\mu i}$ since the remaining $L^{\pm ^{\prime }i}$ are obtained from these
by SO$\left( d-n-1,2\right) $ rotations. The transformations generated by $%
L^{\mu i}$ are given by evaluating the Poisson brackets $\delta \mathbf{u}%
^i=\left\{ \varepsilon _{\nu j}L^{\nu j},\mathbf{u}^i\right\} ,\,\,\delta
x^\mu =\left\{ \varepsilon _{\nu j}L^{\nu j},x^\mu \right\} $

\begin{equation}
\delta \mathbf{u}^i=2\varepsilon ^{\nu j}x_\nu \mathbf{u}_j\mathbf{\,u}%
^i-x_\nu \varepsilon ^{\nu i}\mathbf{u}^2,\quad \delta x^\mu =\varepsilon
^{\mu j}\frac{\mathbf{u}_j}{\mathbf{u}^2}.
\end{equation}
The Lagrangian transforms as follows 
\[
\delta \left( \frac{\mathbf{\dot{u}}^2}{\mathbf{u}^2}+\mathbf{u}^2\dot{x}%
^2\right) =\left( 2\varepsilon _{\mu i}x^\mu \mathbf{u}^i\right) \left( 
\frac{\mathbf{\dot{u}}^2}{\mathbf{u}^2}+\mathbf{u}^2\dot{x}^2\right) . 
\]
This is equivalent to a conformal rescaling of the metric which can be
cancelled by a transformation of the einbein 
\begin{equation}
\delta A^{22}=\left( 2\varepsilon ^{\nu j}x_\nu \mathbf{u}_j\right) \mathbf{%
\,}A^{22}
\end{equation}
Therefore the action for a particle on AdS$_{d-n}\times S^n$ is invariant
under SO$\left( d,2\right) $ for all $n$.

\section{SO(d,2) generators in first quantization}

Since the $AdS_{d-n}\times S^n$ case is of current interest due to the
proposed $AdS-CFT$ duality \cite{maldacena}, we will also discuss the first
quantized theory in that gauge. We will resolve quantum ordering ambiguities
in the generators of SO$\left( d,2\right) $, and then compute the quadratic
Casimir eigenvalue of SO$\left( d,2\right) $ for all values of $n$ at fixed $%
d$, to show that these gauge invariant quantities are independent of $n$ and
are the same as those computed in other gauges, namely $C_2\left( SO\left(
d,2\right) \right) =1-d^2/4$. This confirms the gauge invariant prediction
of two-time-physics, thus verifying its presence.

The full \textit{physical information of the theory is contained in the
gauge invariant }$L^{MN}$. Using the constraints $X^2=P^2=X\cdot P=0$ it is
straightforward to show that all the Casimir operators of SO$\left(
d,2\right) $ vanish at the classical level 
\begin{equation}
Classical:\quad C_n\left( SO\left( d,2\right) \right) =\frac 1{n!}Tr\left(
iL\right) ^n=0,
\end{equation}

In the first quantized theory the $C_n\left( SO\left( d,2\right) \right) $
are not zero after taking quantum ordering into account. Since the $L^{MN}$
are gauge invariant we must find the same eigenvalues in any gauge. First
consider the SO$\left( d,2\right) $ covariant quantization without choosing
any gauges, as treated in \cite{dualconf}. In this case all components of $%
\left( X^M,P^M\right) $ are independent canonical degrees of freedom and the
first class constraints are applied on the states. The constraints form the
Sp$\left( 2,R\right) $ algebra. The states are labelled simultaneously by
the Casimirs of Sp$\left( 2,R\right) $ as well as the Casimirs of SO$\left(
d,2\right) $ since these groups commute $|C_2\left( Sp\left( 2,R\right)
\right) ,C_n\left( SO\left( d,2\right) \right) >$. We need to find their
eigenvalues for physical states. The following relations are proven by
writing out all the Casimir operators in terms of $X,P$. First, all Casimir
eigenvalues $C_n\left( SO\left( d,2\right) \right) $ are rewritten in terms
of $C_2\left( SO\left( d,2\right) \right) $ and $d$. For example $C_3\left(
SO\left( d,2\right) \right) =\frac d{3!}C_2\left( SO\left( d,2\right)
\right) $. Second, the quadratic Casimir of Sp$\left( 2,R\right) $ is
related to the quadratic Casimir of SO$\left( d,2\right) .$ Third, since
physical states are gauge invariant, the quadratic Casimir of Sp$\left(
2,R\right) $ must vanish in the physical sector. The last condition fixes
all the Casimir eigenvalues for SO$\left( d,2\right) $ to unique non-zero
values in terms of $d$. Therefore the quantum system can exist only in a
unique unitary representation of $SO\left( d,2\right) $ characterized by 
\begin{equation}
Quantum:\quad C_2(Sp\left( 2\right) )=0,\quad \left\{ 
\begin{array}{l}
C_2\left( SO\left( d,2\right) \right) =1-\frac{d^2}4, \\ 
C_3\left( SO\left( d,2\right) \right) =\frac d{3!}\left( 1-\frac{d^2}4\right)
\\ 
\cdots
\end{array}
\right. \quad  \label{quantum}
\end{equation}

The first quantization of the theory in several one-time physics gauges
(massless relativistic particle, H-atom, harmonic oscillator, and all of
these with spin) has already been performed elsewhere \cite{dualconf}-\cite
{dualsusy}, and the correct value of the Casimirs (which change with spin)
have been obtained, in agreement with the prediction. Hence for these
diverse systems the Hilbert space corresponds to the same unique
representation of SO$\left( d,2\right) $. This establishes a Sp$\left(
2,R\right) $ duality among these systems at the quantum level. This may be
considered a successful test of the unification in the form of two-time
physics at the quantum level.

Now we consider the first quantized theory in the $AdS_{d-n}\times S^n$
gauge. We want to find the correct operator ordering of the generators in
the quantum theory and then compute the quadratic Casimir eigenvalue. We
must find that $C_2\left( SO\left( d,2\right) \right) =1-\frac{d^2}4$ since
this is the prediction of the gauge invariant two-time physics. Confirming
this result is tantamount to the presence of two time physics in the
one-time quantum theory of the $AdS_{d-n}\times S^n$ particle, and to
establishing that the Hilbert space is the same as the other cases already
mentioned.

With operator ordering taken into account the quantum generators have the
form 
\begin{equation}
L^{MN}=\left| \mathbf{u}\right| ^{-d/2+n+2}L_0^{MN}\,\left| \mathbf{u}%
\right| ^{d/2-n-2}.  \label{ulmnu}
\end{equation}
Evidently the factors of $\left| \mathbf{u}\right| ^{\pm \left(
d/2-n-2\right) }$ drop out in the classical theory, but they are essential
for the correct symmetry generators in the quantum theory as proved in the
last section (see eq.(\ref{quantumsymm})). The (non-unitary looking)
similarity transformation with the $\left| \mathbf{u}\right| ^{d/2-n-2}$
will be explained in the next section. This transformation is required for
hermiticity of the generators according to a scalar product defined in eq.(%
\ref{scalarprod}) that is appropriate for an AdS covariant quantization
scheme. The $L^{MN}$ are hermitian in the physical Hilbert space provided
the $L_0^{MN}$ are the following operator ordered versions of the classical
generators (\ref{lmnclass1}-\ref{lmnclass5}) 
\begin{eqnarray}
L_0^{+^{\prime }-^{\prime }} &=&\frac 12\left( x\cdot p+p\cdot x-\mathbf{%
u\cdot k}-\mathbf{k\cdot u}\right) +i  \label{lmnq1} \\
L_0^{+^{\prime }\mu } &=&p^\mu ,\quad L_0^{+^{\prime }i}=\mathbf{V}^i\left( 
\mathbf{u,k}\right) \\
L_0^{-^{\prime }\mu } &=&\frac 12x^\nu p^\mu x_\nu -\frac 12x\cdot px^\mu
-\frac 12x^\mu p\cdot x-ix^\mu \\
&&+\frac{p^\mu }{2\mathbf{u}^2}+\frac 12\left( \mathbf{u\cdot k+k\cdot u}%
\right) x^\mu \\
L_0^{-^{\prime }i} &=&\frac 12\mathbf{k}^i+\frac{x^2}2\mathbf{V}^i\left( 
\mathbf{u,k}\right) -\frac 12\left( x\cdot p+p\cdot x+2i\right) \frac{%
\mathbf{u}^i}{\mathbf{u}^2} \\
L_0^{\mu \nu } &=&x^\mu p^\nu -x^\nu p^\mu ,\quad L_0^{ij}=\mathbf{u}^i%
\mathbf{k}^j-\mathbf{u}^j\mathbf{k}^i \\
L_0^{\mu i} &=&x^\mu \mathbf{\,V}^i\left( \mathbf{u,k}\right) -p^\mu \frac{%
\mathbf{u}^i}{\mathbf{u}^2}  \label{lmnq7}
\end{eqnarray}
$\mathbf{V}^i\left( \mathbf{u,k}\right) $ is the operator ordered version of
the classical expression $\mathbf{u}^2\mathbf{k}^i\,\mathbf{\,}-2\,\mathbf{%
\,k\cdot u}\,\mathbf{u}^i$ 
\begin{eqnarray}
\mathbf{V}^i\left( \mathbf{u,k}\right) &=&\mathbf{u}^k\mathbf{k}^i\mathbf{u}%
^k-\mathbf{u}^i\mathbf{k\cdot u}\,-\mathbf{u\cdot k}\,\mathbf{u}^i \\
&=&\mathbf{u}^2\mathbf{k}^i-\mathbf{u}^i\left( \mathbf{u\cdot k+k\cdot u}%
\right) \, \\
&=&\mathbf{k}^i\mathbf{u}^2-\left( \mathbf{u\cdot k+k\cdot u}\right) \mathbf{%
u}^i\,
\end{eqnarray}
Some of its interesting properties are 
\begin{eqnarray}
\left[ \frac{\mathbf{u}^i}{\mathbf{u}^2},\mathbf{V}^j\right] &=&i\delta
^{ij},\quad \left[ \mathbf{V}^i,\mathbf{V}^j\right] =0,\quad \mathbf{V}^2=%
\mathbf{u}^2\mathbf{k}^2\mathbf{u}^2  \label{Vi} \\
\left[ \mathbf{k}^i,\mathbf{V}^j\right] &=&-2iL^{ij}+i\delta ^{ij}\left( 
\mathbf{u\cdot k+k\cdot u}\right) \\
\left[ \mathbf{u}^i,\mathbf{V}^j\right] &=&i\delta ^{ij}\left( \mathbf{u}^2-2%
\mathbf{u}^i\mathbf{u}^j\right) ,\quad \left[ \left| \mathbf{u}\right| ,%
\mathbf{V}^j\right] =-i\left| \mathbf{u}\right| \mathbf{u}^j
\end{eqnarray}
These may be used to verify the closure of the algebra at the quantum level%
\footnote{%
The following change of variables $\mathbf{r}^i\mathbf{=}\frac{\mathbf{u}^i}{%
\mathbf{u}^2},\,\mathbf{p}^i=\mathbf{V}^i\left( \mathbf{u,k}\right) $ is a
canonical transformation at the quantum level. With this substitution one
may notice that the generators of SO$\left( d,2\right) $ take the same form
we found in \cite{dualconf} for the free massless relativistic particle.
Hence the computation of the Casimir operator is easily explained. This may
also shed light on the overall structure of the generators, and helps
explain the anomaly terms $i$ in $L^{+^{\prime }-^{\prime }}$, $-ix^\mu $ in 
$L^{-^{\prime }\mu }$, and $i\mathbf{u}^i/\mathbf{u}^2$ in $L_0^{-^{\prime
}i}$, as due to hermiticity in Lorentz covariant quantization in flat space.
The AdS covariant quantization introduces the further anomaly terms
generated by the similarity transformation $\left| \mathbf{u}\right| ^{\pm
\left( d/2-n-2\right) }$ given in (\ref{ulmnu}).}. Since $\left| \mathbf{u}%
\right| ^{\pm \left( d/2-n-2\right) }$ is a similarity transformation, the
commutation relations are the same for $L^{MN}$ or $L_0^{MN}$ at the quantum
level.

The quadratic Casimir may now be computed for $L_0^{MN}$ or $L^{MN}$. All
operators cancel and it reduces to a pure number independent of $n$ for all
AdS$_{d-n}\times S^n$ 
\begin{equation}
C_2\left( SO\left( d,2\right) \right) =1-d^2/4.
\end{equation}
This is the correct value imposed by the overall structure of two-time
physics as given in eq.(\ref{quantum}). The similarity transformation of eq.(%
\ref{ulmnu}) cannot change the Casimirs 
\begin{equation}
C_n\left( L^{MN}\right) =\left| \mathbf{u}\right| ^{-d/2+n+2}C_n\left(
L_0^{MN}\right) \,\left| \mathbf{u}\right| ^{d/2-n-2}=C_n\,\,,
\end{equation}
since they are pure numbers.

\section{Field Theory in AdS$_{d-n}\times S^n$ background}

As seen from the action (\ref{adsaction1}) the equation of motion for $A^{22}
$ generates the classical constraint 
\begin{equation}
P^2=\left( \frac{p^2}{\mathbf{u}^2}+\mathbf{u}^2\mathbf{\,k}^2\right)
=G^{mn}p_mp_n=0,\quad p_m\equiv \left( p_\mu ,\mathbf{k}_i\right) .
\end{equation}
In the quantum theory the constraint is applied on the Hilbert space to find
the physical states which are annihilated by it 
\begin{equation}
\,\left( :G^{mn}p_mp_n:\right) |\phi >=0.
\end{equation}
The columns $\left( :\right) $ indicate that the operator form of the
constraint must first be defined by resolving ambiguities in the ordering of
the operators. This must be done in such a way as to preserve the SO$\left(
d,2\right) $ symmetry of the system.

One possible ordering follows from the definition of Laplacian in General
Relativity. This is guaranteed to preserve the symmetries SO$\left(
d-n-1,2\right) \times S\left( n+1\right) $ of the background AdS$%
_{d-n}\times S^n$, so it is a good starting point. In configuration space
the constraint on the wavefunction $\phi \left( x^\mu ,\mathbf{u}^i\right)
=<x^\mu ,\mathbf{u}^i|\phi >$ takes the form of the Laplace equation 
\begin{equation}
\partial _m\left( \sqrt{-G}G^{mn}\partial _n\phi \left( x^\mu ,\mathbf{u}%
^i\right) \right) =0.
\end{equation}
This follows from the effective action $S_{eff}=\frac 12\int d^dX\,\sqrt{-G}%
\left( G^{mn}\partial _m\phi \,\partial _n\phi \right) $. Using 
\begin{eqnarray}
G_{mn} &=&\left( 
\begin{array}{ll}
u^2\eta _{\mu \nu } & 0 \\ 
0 & \frac 1{u^2}\delta _{ij}
\end{array}
\right) ,\quad \sqrt{-G}=u^{d-2n-2},\quad  \\
G^{mn} &=&\left( 
\begin{array}{ll}
\frac 1{u^2}\eta ^{\mu \nu } & 0 \\ 
0 & u^2\delta ^{ij}
\end{array}
\right) ,
\end{eqnarray}
we find the effective action 
\begin{eqnarray}
S_{eff}^0\left( \phi \right)  &=&\frac 12\int d^{d-n-1}x\,d^{n+1}u\,\,\left(
u^{d-2n-4}\partial _\mu \phi ^{*}\,\partial ^\mu \phi +u^{d-2n}\partial
_i\phi ^{*}\,\partial _i\phi \right)  \\
&=&\frac 12<\phi |\left( \left| \mathbf{u}\right| ^{d-2n-4}p^2+\mathbf{k}%
^i\left| \mathbf{u}\right| ^{d-2n}\mathbf{k}^i\right) |\phi >
\end{eqnarray}
The norm of the state is not $<\phi |\phi >=\int \phi ^{*}\phi $, but rather
it is defined by the scalar product ($\phi ,\phi )=\int d\Sigma _mJ^m$,
using the conserved probability current $J^m=\sqrt{-G}G^{mn}\left( \phi
^{*}\,i\partial _n\phi -i\partial _n\phi ^{*}\,\phi \right) $, by
integrating over a spacelike surface, such as fixed time 
\begin{eqnarray}
(\phi ,\phi ) &=&\int_{x^0=fixed}\,\sqrt{-G}G^{0n}\left( \phi
^{*}\,i\partial _n\phi -i\partial _n\phi ^{*}\,\phi \right)  \\
&=&\int_{x^0=fixed}\left( d^{d-n-2}x\right) \left( d^{n+1}\mathbf{u}\right)
\,\left| \mathbf{u}\right| ^{d-2n-4}\left( \phi ^{*}\,i\partial _0\phi
-i\partial _0\phi ^{*}\,\phi \right) .  \label{scalarprod}
\end{eqnarray}
The adjoint of an operator and its hermiticity in the physical Hilbert space
must be defined relative to this scalar product. This approach defines the 
\textit{AdS-covariant quantization scheme} consistent with field theory. The
operators $L^{MN}$ defined in the previous section are hermitian according
to the scalar product in this quantization scheme. This explains the reason
for the similarity transformation $\left| \mathbf{u}\right| ^{\pm \left(
d/2-n-2\right) }$ and the other insertions of $i$ in eqs.(\ref{ulmnu},\ref
{lmnq1}-\ref{lmnq7}). The analog of this approach for Lorentz covariant
quantization of the relativistic particle in flat space, with and without
spin, was discussed in \cite{dualconf},\cite{dualsusy}.

The Laplace equation may be written in operator form $\hat{S}_0|\phi >=0$,
where 
\begin{equation}
\hat{S}_0=\left| \mathbf{u}\right| ^{d-2n-4}p^2+\mathbf{k}^i\left| \mathbf{u}%
\right| ^{d-2n}\mathbf{k}^i.
\end{equation}
This is just the constraint condition with a particular order of operators.
Thus, general covariance imposes a particular order. To check the symmetries
of the effective field theory action $S_{eff}\left( \phi \right) $ for this
order of operators we transform the wavefunction 
\begin{equation}
\delta |\phi >=-\frac i2\varepsilon _{MN}L^{MN}|\phi >,\quad \delta <\phi
|=\frac i2\varepsilon _{MN}<\phi |\left( L^{MN}\right) ^{\dagger }.
\end{equation}
Then the transformation of the action $\delta S_{eff}^0\left( \phi \right) $
can be written in the form 
\begin{equation}
\delta S_{eff}^0\left( \phi \right) =\frac i2\varepsilon _{MN}<\phi |\left[
\left( L^{MN}\right) ^{\dagger }\hat{S}_0-\hat{S}_0L^{MN}\right] |\phi >.
\end{equation}
Note that $\left( L^{MN}\right) ^{\dagger }$ is the naive hermitian
conjugation\footnote{%
Because of the $i$ insertions and the factors of $\left| \mathbf{u}\right|
^{\pm \left( d/2-n-2\right) }$ in eqs.(\ref{ulmnu},\ref{lmnq1}-\ref{lmnq7}) $%
\left( L^{MN}\right) ^{\dagger }$ may not be equal to $L^{MN}$. This is of
no concern since the $L^{MN}$ are hermitian in the correct sense defined
above, not in the naive sense.} (using $x,p,\mathbf{u},\mathbf{k}$ that are
naively hermitian). Now we can verify that the generators $L^{ij}$ and $%
\left( L^{+^{\prime }-^{\prime }},L^{+^{\prime }\mu },L^{-^{\prime }\mu
},L^{\mu \nu }\right) $ are indeed symmetries of the effective action as
expected in the General Relativity formalism. Indeed we find explicitly $%
\delta S_{eff}^0\left( \phi \right) =0$, confirming the SO$\left(
d-n-1,2\right) \times SO\left( n+1\right) $ invariance of the action and of
the quantization procedure.

Next we turn to the remaining generators of SO$\left( d,2\right) $, $%
L^{+^{\prime }i},L^{-^{\prime }i},L^{\mu i}$. We find that the ordering of
operators given by $\hat{S}_0$ introduces quantum anomalies that break the
bigger symmetry SO$\left( d,2\right) $ for the generic AdS$_{d-n}\times S^n$
background in field theory. There are exceptions for the cases AdS$_2\times
S^0$ and AdS$_n\times S^n$ (i.e. $d=2n$) for which the anomaly is zero and
the full symmetry is active. On the other hand it is also possible to
improve the effective action by adding the following anomaly term to the
action in such a way as to preserve the full SO$\left( d,2\right) $ for all $%
d,n$%
\begin{eqnarray}
\hat{S} &=&\hat{S}_0-\frac 14\left( d-2\right) \left( d-2n\right) \left| 
\mathbf{u}\right| ^{d-2n-2}, \\
S_{eff}\left( \phi \right)  &=&\frac 12<\phi |\hat{S}|\phi >=S_{eff}^0\left(
\phi \right) +S_{eff}^1\left( \phi \right) .
\end{eqnarray}
The anomaly $S_{eff}^1\left( \phi \right) $ results from a different
ordering of the operators and may be seen as a potential term (no momenta)
added on to the kinetic term defined by General Relativity. Actually it is
just a mass term in the field theory formalism, $S_{eff}^1\left( \phi
\right) =-\frac 12m^2\int \sqrt{-G}\phi ^{*}\phi $, with the quantized mass
(in units of the S$^n$ radius$^{-2}$ that was set to $R=1$) 
\begin{equation}
m^2=\frac 14\left( d-2\right) \left( d-2n\right) .
\end{equation}
Of course, the mass term is invariant separately under the subgroup SO$%
\left( n+1\right) \times SO\left( d-n-1,2\right) $. On the other hand, the
total action is invariant under the full SO$\left( d,2\right) $ thanks to
the relations 
\begin{equation}
\left( L^{MN}\right) ^{\dagger }\hat{S}-\hat{S}L^{MN}=0  \label{quantumsymm}
\end{equation}
that are satisfied just for the special value of the mass, and the precise
ordering of operators in $L^{MN}$ as given in eqs.(\ref{ulmnu},\ref{lmnq1}-%
\ref{lmnq7}). For the special cases AdS$_2\times S^0$ and AdS$_n\times S^n$
(i.e. $d-n=n$) the mass vanishes.

In recent literature the cases of AdS$_2,\,$AdS$_3\times S^3$ and AdS$%
_5\times S^5$ have been investigated in the context of the AdS/CFT
correspondance \cite{maldacena}-\cite{muratminic}. These are among the cases
that, according to our results, have higher symmetries $SO\left( 2,2\right) $%
, SO$\left( 6,2\right) $ and SO$\left( 10,2\right) $ respectively, with
vanishing mass term. The higher symmetry may be of interest in future
investigations.

We have shown that in order to be consistent with two-time physics the
quantum theory must be carefully constructed. The formalism puts constraints
that are non-trivial. One of the signatures of two-time physics is the SO$%
\left( d,2\right) $ symmetry realized in a unique unitary representation
with special values of the Casimir operators. This is a unifying aspect
since it connects diverse one-time physics systems in the same quantum
representation of SO$\left( d,2\right) $. Furthermore, our work establishes
a quantum duality for AdS$_{d-n}\times S^n$ for all $n$ among themselves, as
well as with all other one-time physics systems that we derived before \cite
{dualconf}-\cite{dualicmp} from the same action.

\end{document}